# Highly Sensitive and Self Powered Ultraviolet Photo Detector based on ZnO Nanorods Coated with TiO$_2$

Shashi Pandey[1], Alok Shukla[2*], Anurag Tripathi[1]

[1]*Department of Electrical Engineering IET Lucknow, Uttar Pradesh 226021, India*

[2]*Department of Physics, Indian Institute of Technology Bombay, Powai, Mumbai 400076, India*

**Abstract**

Nanorods (NRs) of crystalline ZnO coated with thin layers of TiO$_2$ (ZnO@TiO$_2$) were fabricated with the help of spin coating technique followed by hydrothermal method. Scanning electron microscopy (SEM) and X-ray diffraction analysis confirms the morphology and structural stability of as-prepared NRs. The optical band gaps of the NRs were estimated, and a clear blue-shift towards the UV region has been detected. When UV light falls on as-prepared device (i.e., in the "ON" state), a significant increase in photocurrent ($I_{UV}$) at zero voltage supply was observed from 6 µA to 17 µA, while in the "OFF" state, the dark current ($I_{dark}$), increases from 0.08 µA to 0.6 µA with ZnO@TiO$_2$ NRs as compared to bare ZnO NRs respectively. Responsivity and detectivity of TiO$_2$ coated ZnO NRs based device found maximum in UV region unlike bare ZnO NRs. Enhanced photocurrent achieved by the growth of TiO$_2$ layers on ZnO NRs is 250 µA as compared to bare ZnO NRs for which it is 35 µA at 10V voltage supply under the ultraviolet irradiation (illumination intensity of 1 mW/cm$^2$). Furthermore, theoretical calculations have been performed using the first-principles density-functional theory to understand the effects of heterostructure NRs on the electronic and optical properties of TiO$_2$ coated ZnO.

*Email: 2512@ietlucknow.ac.in, shukla@phy.iitb.ac.in, anurag.tripathi@ietlucknow.ac.in*

*\*corresponding author*

**Introduction**

Ultraviolet photo detection has many applications in the field of optical imaging, optoelectronic circuits, military surveillance, air quality monitoring, and even in space communication [1]–[4]. Conventional photo detectors based on materials such as silicon, germanium, gallium arsenide, and silicon carbide, etc. become expensive because they require high temperature conditions for device fabrication, as well as visible light filters. Therefore, the invention of fast, sensitive, direct UV detectors, that are easy to synthesize, is important. For high UV sensitivity, we need materials that are transparent in the lower-energy region of the spectrum, and have strong optical response in the UV region. Devices based on several semiconducting oxides are very useful for the purpose because of their environment friendly nature, non-toxic character, large band-gap leading to high UV sensitivity, low cost, and excellent thermal stability. The sensitivity to the UV light in these materials can be further tuned by constructing nanostructures of different shapes and sizes, and also by managing oxygen adsorption and native defects [5]–[7].

In this work we have explored the optical response of nano-rods (NRs) of a transition metal oxide, namely, crystalline ZnO, coated with thin layers of $TiO_2$ (ZnO@$TiO_2$). Pure ZnO is not a good candidate for such devices because it has inherent deep levels and surface defects [8] which enhance undesirable photocurrent leading to low sensitivity and switching stability during device operation. Recently, a few groups studied the synthesis and optical response of $TiO_2$ and ZnO-based nanocomposites, with the purpose of demonstrating improved device performance [9], [10], as compared to bare ZnO nanostructures. $TiO_2$ has a wide range of applications [8], [11]–[13] such as in photo detectors, gas sensors, and thermal catalysis. In several studies, enhancement in UV photo detection properties of ZnO, by changing its morphology, was pursued [14]–[19]. On the other hand, many works have suggested that doping of different materials in ZnO can improve its optoelectronic and photoconductive properties [3], [20]–[22]. Because of its excellent thermal and photochemical stability, $TiO_2$ has been investigated as a candidate for enhancing the optoelectronic properties of nanocomposite devices [23]. In order to increase the sensitivity of photodetectors, researchers have focused more on the coupling of semiconductors, particularly a heterojunction based on two types of semiconductors with differing energy band structures [24]. As a result of the coupling of various energy level structures, $TiO_2$ and ZnO are strong candidates to produce a heterostructure with improved characteristics compared to those of either material alone. Wang et al [24] reported that ZnO@TiO$_2$ nanostructures also play an important role in dye-sensitized solar cells applications. ZnO is a well-known n-type semiconductor due to the presence of native defects [25]–[27], as a result of which electrons (i.e., its majority charge carriers) already exist in its conduction band at the room temperature[28]. As a result of this, while measuring the photocurrent; one sees a drop in the photocurrent in the UV "OFF" state of the bare ZnO samples. However, in the UV "ON" state, the photocurrent increases because of the photon-induced promotion of the valence band electrons to the conduction band in bare ZnO[28], [29]. As far as ZnO@TiO$_2$ NRs are concerned, because of the high reactivity of anatase TiO$_2$, a significant O$_2$ desorption takes place, thus freeing up electrons, leading to a large rise in the photocurrent in the UV "ON" state, as compared to the bare ZnO NRs [16], [18], [19], [24]. Hence, this work will improve the understanding of heterostructures based on semiconducting oxides, which leads to enhance the optoelectronic properties of devices.

**Experimental Procedure**

*A. Synthesis and Characterization of ZnO@TiO$_2$ Nano rods*

Firstly, cleaning of ITO substrate was done using acetone and methanol or IPA. ITO substrate cleaned with firstly under acetone and then put in methanol quickly before it dries after taking out from acetone. Then the mixture of zinc acetate dehydrates and ethanolamines (with ratio 1:2) were dissolved in 25 mL 2-methoxy ethanol. Next, the prepared solution was stirred at 50 $^0$C for 1 hour, and then spin-coated with 2500 rpm for 1 min onto indium-tin oxide (ITO) glass substrates. The resultant sample was further heated at 350 $^0$C for 6 hours to obtain ZnO NRs on the ITO substrate. Next, the prepared NRs of ZnO were placed inside an autoclave, which contained diethylenetriamine, isopropyl alcohol, and a titanium (IV) isopropoxide. Autoclave, containing the mixture, was given a heat treatment (200 $^0$C for 12 hours), after which the as-prepared sample was smoothly rinsed with ethanol and acetone at the room temperature. Finally, the rinsed sample was annealed at 450 $^0$C for 6 hours to confirm the deposition (i.e. coating) of crystalline phases of TiO$_2$ on ZnO NRs. For structural and morphological characterizations, Zeiss Supra-55 field emission scanning electron microscope (FESEM) was used. In order to examine the structural phase purity of the prepared samples, the powder x-ray diffraction (XRD) experiments were carried out on Bruker D8 diffractometer equipped with Cu target having LYNXEYE detector. The high temperature x-ray diffraction measurements were performed to confirm the structural phases of the prepared samples. The optical band gap of prepared sample has been measured using diffuse reflectivity measurements. These measurements have been performed in the 200 nm to 800 nm wavelength range using Perkin Elmer LAMBDA 950 UV-Vis-NIR Spectrophotometer. For device fabrication Ag (thickness 50 nm) has been deposited as an electrode using thermal evaporation. Shadow masking was used to keep the width to 1 mm and the distance between the Ag electrodes (channel length) to 100 µm. Electrical characterizations were performed by Keithley 2612A source meter.

*B. Computational details*

The first-principles calculations to support the experimental results were performed within the framework of plane-wave density-functional theory (PW-DFT) [30], [31] as implemented in Vienna ab-initio simulation package (VASP)[32], [33]. To match our calculated bandgap with the experimental results, the generalized gradient approximation coupled with the Hubbard U (GGA+U) level of theory was used [34], [35]. We performed calculations both on the bare ZnO NRs, and the ZnO@$TiO_2$ NRs, and the super cells used for the two systems are shown in Fig. 1. To model the bare NRs, we used a one-dimensional super cell of dimension 12x1x1 (see Fig. 1(a)) of bulk ZnO, while for simulating ZnO@$TiO_2$ NRs, we added one-unit cell of the anatase phase of $TiO_2$ containing twelve atoms (4 Ti and 8 O) on one of the ends of the ZnO super cell. Furthermore, to model the finite size of the NRs, 10 Å of vacuum was used in two directions, as shown in Fig. 1. One can ask as to why did we model ZnO@$TiO_2$ NRs by attaching a unit cell of $TiO_2$ only on one tip of the bare NRs, and not on the second tip and other places? The reason behind this choice is that the SEM images presented in the next section clearly show that the ZnO@$TiO_2$ NRs have $TiO_2$ attached only on one tip of the bare NRs, and nowhere else. As far as the optimized geometries of the two types of NRs presented in Fig. 1 are concerned: (a) bare NRs were found to be 39.28 Å long, with a diameter of 5.20 Å (Fig. 1(a)), and (b) the total length of ZnO@TiO$_2$ NRs is approximately 43.01 Å, with the diameter on the ZnO side of around 5.20 Å, and on the TiO$_2$ side 3.73 Å (Fig. 1(b)).

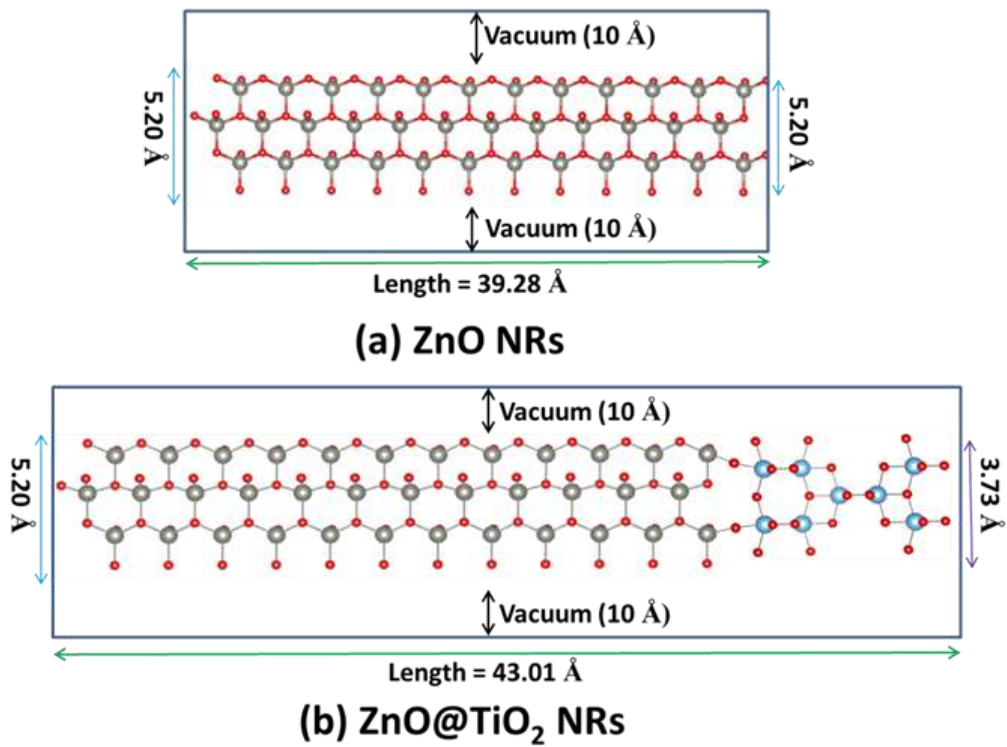

**FIGURE 1:** *Structural Images of (a) bare ZnO having 12x1x1 supercell and (b) composite of ZnO having supercell 12x1x1 along with TiO$_2$ with single unit cell. Blue and red color indicates Zn, Ti and O atoms respectively.*

**Results and Discussion**

Figure 2(a) shows the SEM image of the as-prepared ZnO@TiO$_2$ on ITO glass substrate. Firstly, the ZnO NRs have been deposited on ITO glass substrate using spin coating method (see Figure 2(b)). Figure 2(b) shows the vertical orientation of ZnO NRs (top view). Deposition of TiO$_2$ on ZnO NRs was done using the hydrothermal process, and as-prepared TiO$_2$ nanocomposites on ZnO[37] are also shown in Figure 2(a) and 2(b). The average length and diameter of the NRs were around 1.80 μm and 150 nm with particle size 3-23.2 nm. From the top view of SEM image, TiO$_2$ has been deposited only on the exposed tips of ZnO NRs (see figure 2(b)), The nanostructures (NRs) are typically about 1 μm in length and 150–200 nm in diameter. Hence, the SEM analysis confirms the fabrication of ZnO@TiO$_2$ NRs.

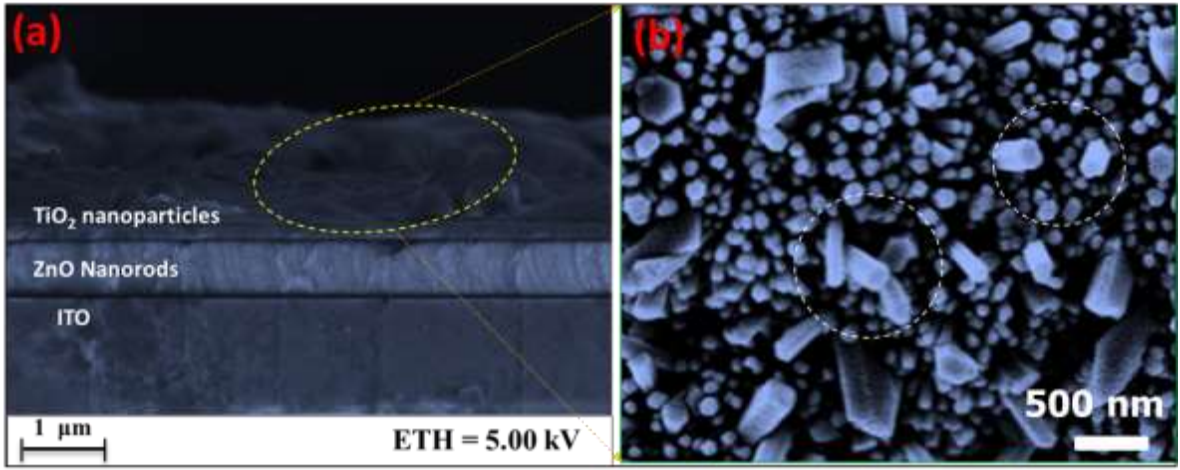

**FIGURE 2:** *FESEM images of (a) deposited TiO$_2$ on ZnO NRs, (b) shows enlarged view (yellow color) of deposited TiO$_2$ nanoparticles on ZnO NRs.*

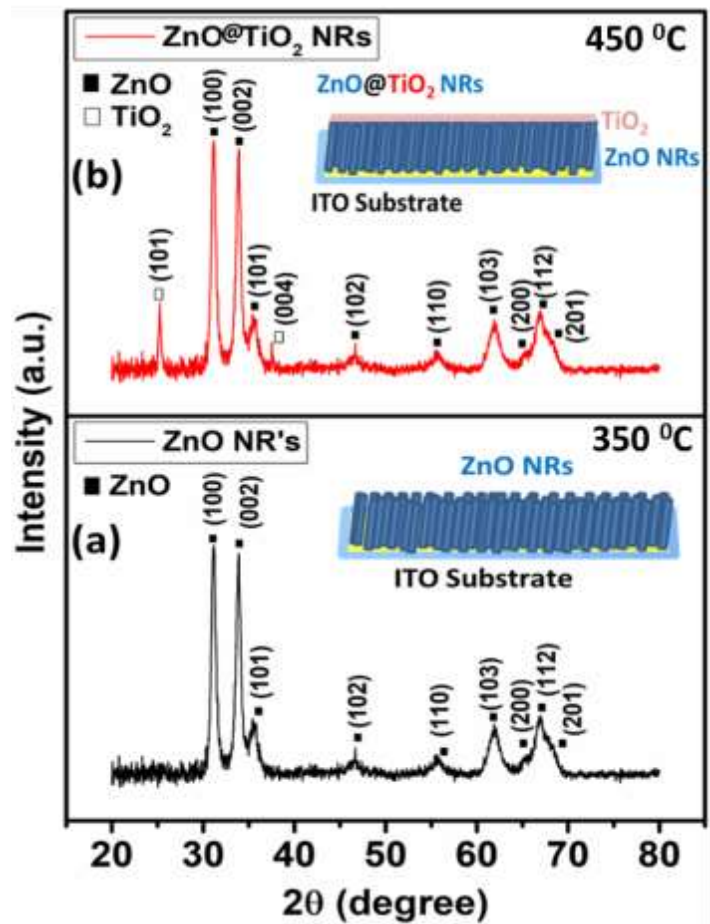

**FIGURE 3:** *X-Ray diffraction patterns of as-prepared ZnO and ZnO@TiO$_2$ NRs.*

To probe the structural stability of our prepared ZnO@TiO₂ nanocomposites, we have performed X-ray diffraction analysis (see Figure-3). X-ray diffraction patterns of the as-prepared ZnO NRs have been shown in figure-3(a) and indexing of all the diffraction peaks confirms the hexagonal wurtzite structure of ZnO[38], [39]. The x-ray diffraction pattern of ZnO@TiO₂ is shown in Figure 3(b) and additional peaks corresponding to (101) and (004) planes of the anatase phase of TiO₂ are clearly observed. These coupled with the SEM images further confirm the formation of the ZnO@TiO₂ composite NRs, with the TiO₂ deposited on the tips. We have probed the optical absorption for bare ZnO NRs and TiO₂ coated ZnO NRs using diffuse reflectance spectroscopy. The spectra obtained from DRS is converted into equivalent absorption spectra through Kubelka–Munk equation [8,13],

$$F(R_\infty) = \frac{(1-R_\infty)^2}{2R_\infty}, \qquad (i)$$

Where $F(R_\infty)$ is the Kubelka–Munk function, $R_\infty = R_{sample}/R_{standard}$. The Kubelka–Munk function can be related (proportional) to the absorption coefficient (α) as-

$$F(R_\infty) \propto \alpha \propto \frac{(h\nu - E_g)^{1/n}}{h\nu}, \qquad (ii)$$

In order to calculate the $E_g$, the obtained absorption coefficient is converted in to Tauc equation [13] and plotted in figure 4.

$$(\alpha h\nu)^n = A(h\nu - E_g), \qquad (iii)$$

Here in equation (iii) 'n' has the value of 2 for direct bandgap transitions (used for ZnO bare NRs), while n is equal to 1/2 for an indirect transition (used for ZnO@TiO₂ NRs) [13]. To see the change in the optical band gap of deposited ZnO[8], [16], [17], and ZnO@TiO₂ NRs, optical absorption measurements have been carried out in the range 300 nm to 900 nm. It is well known that bulk ZnO has a band gap of 3.3 eV, while our measured value of the optical gap of ZnO NRs is 3.46 (see Fig. 4). Clearly, band gap increases in the NR phase, as compared to the bulk [40], [41]. Our measured optical gap of as prepared ZnO@TiO₂ NRs further increases to 3.56 eV (see Fig. 4), which is closer to the UV range. This increase can clearly be attributed to the presence of TiO₂ in the sample.

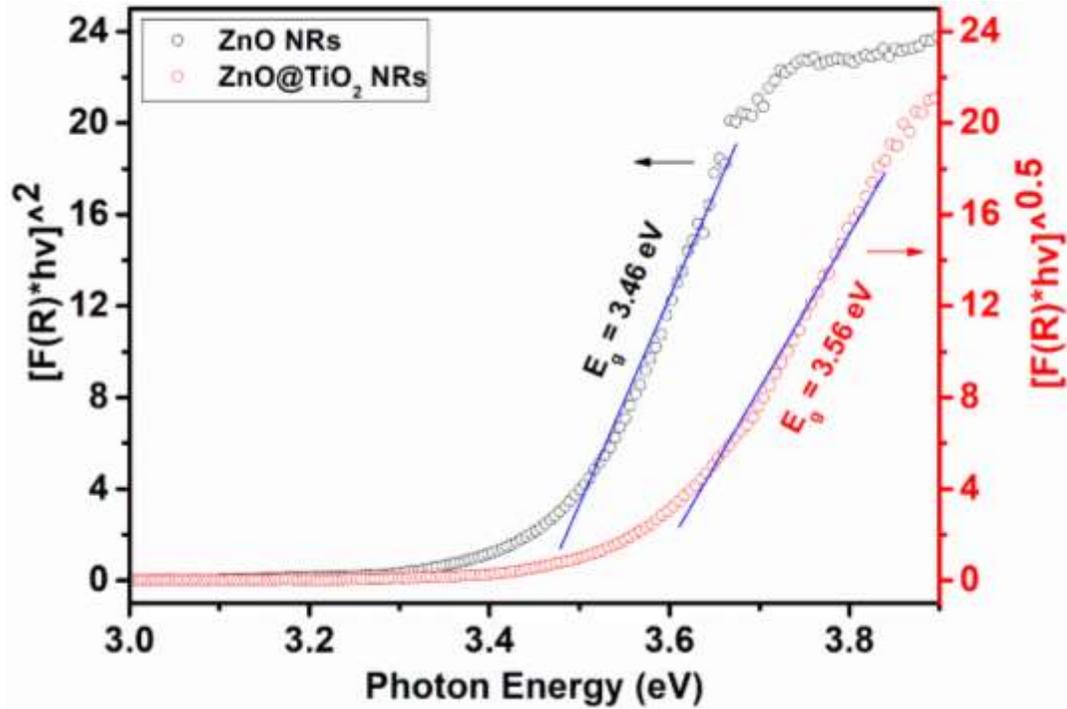

**FIGURE 4:** *Measured optical absorption spectra of ZnO and ZnO@TiO$_2$ NRs.*

To verify the experimental band gap of nanocomposites of ZnO and TiO$_2$, systematic first-principles DFT calculations of total density of states (TDOS)[42] and optical absorption spectra have been performed. Figure 5 show TDOS of bare ZnO NRs with an electronic band gap of 3.46 eV calculated using the GGA+U method [43] with U=4.3 eV (at Zn sites), while the inset shows the optical absorption spectrum obtained from the same calculations. We have performed similar calculations for ZnO@TiO$_2$ NRs, and observed a clear blue shift in band gap as compared to bare ZnO NRs. In Fig. 6 we show TDOS of ZnO@TiO$_2$ NRs with an electronic band gap of 3.58 eV obtained the using the GGA+U calculations with U=4.3 eV at the Zn sites and U=5.6 eV at the Ti sites, along with the computed optical absorption spectrum shown in the inset. Very good quantitative agreement between the theoretically computed band gaps with the experimentally measured ones indicates that our choice of U parameters in the GGA+U calculations is correct (see Figs. 4, 5 and 6).

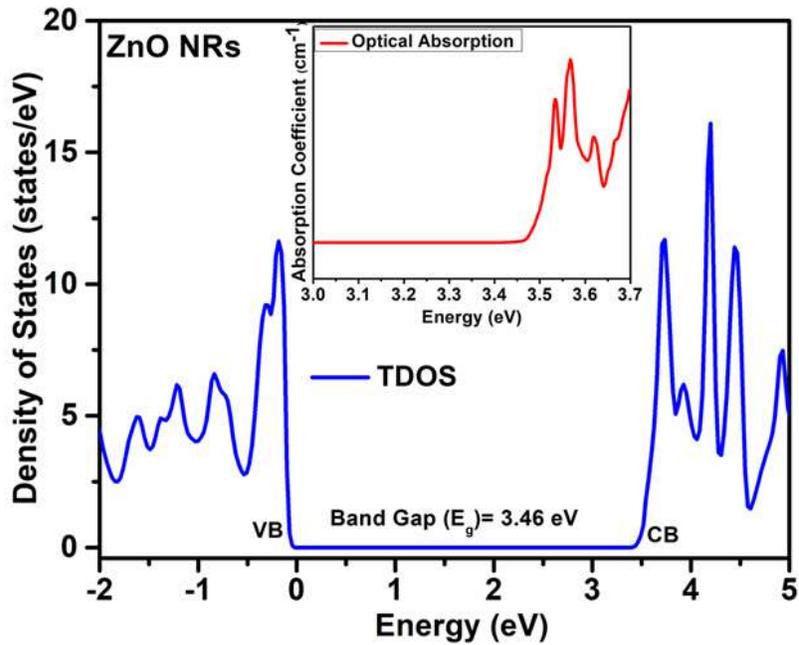

**FIGURE 5:** *TDOS of ZnO NRs exhibits an electronic band gap of 3.46 eV obtained from the GGA+U calculations (U=4.3 eV). The inset shows the calculated optical absorption spectrum of bare ZnO NRs.*

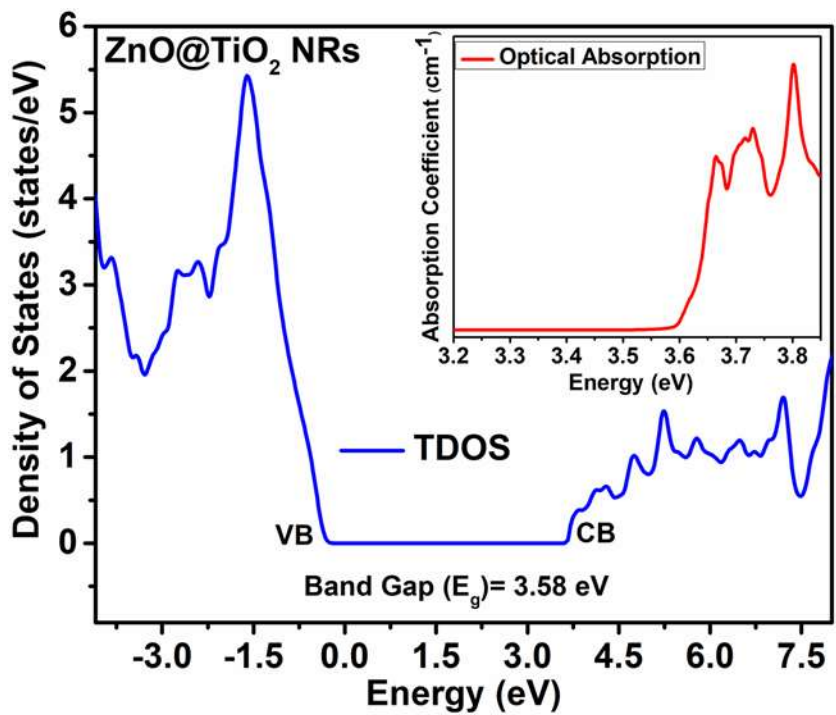

**FIGURE 6:** *TDOS of ZnO@TiO$_2$ NRs shows electonic band gap of 3.58 eV using GGA+U calculations (U=4.3 eV for Zn and U=5.6 eV for Ti). The inset shows simulated optical absorption spectrum of ZnO@TiO$_2$ NRs.*

The measurement setups for photocurrent spectra for ZnO NRs and ZnO@TiO$_2$ are shown in Figs. 7(a) and 7(b), respectively. Before performing the measurements, first we heated the prepared device up to a temperature of 300$^0$C for 8 hours to avoid moisture, and then kept it in dark on vacuum desiccator for several hours to reach the equilibrium condition. I-V characteristics of ZnO NRs and ZnO@TiO$_2$ NRs measured using a UV lamp (with wavelength 365 nm) have been plotted in Figs.-7(c) and 7 (d), and it is clearly observed that with the increase in the voltage, the device current also increases for both cases. When UV is "ON", the device current increases sharply, while for UV "OFF", the increase in the current with the applied voltage is quite negligible. On comparing the I-V characteristics of bare ZnO NRs and ZnO@TiO$_2$ NRs (Figs.-7(c) and 7(d)), we conclude that the TiO$_2$ coated on ZnO NRs enhances the value of the current with applied voltages. This result indicates that bare ZnO NRs as well as ZnO@TiO$_2$ NRs are highly UV sensitive; however, the photoelectric response of ZnO@TiO$_2$ NRs is much more intense. Insets (i) and (ii) of Figs. 7(c) and 7(d) respectively show that there are still small amounts of charge carriers inside the device at V=0, irrespective of whether the UV light is "ON" or "OFF".

To determine the performance of device, responsivity, R = (I$_{UV}$/I$_P$) [19], is an important parameter where, I$_P$ is the incident power and I$_{UV}$ is the maximum current under UV irradiation. From Fig. 8 (a) and 8 (b) it is clearly seen that responsivity of our device is maximum in UV region while it is decreasing in visible region. Detectivity (D = 1/NEP$_B$) [19] has been depicted in Fig. 8(a) and 8(b), and peak value reached at 1.8 × 10$^{14}$ (at ~ 450 nm) and 6.4 × 10$^{14}$ Hz$^{1/2}$/W (at 380 nm) for bare ZnO NRs and TiO$_2$ coated ZnO NRs respectively. It is clearly seen that in case of TiO$_2$ coated ZnO NRs shows enhancement in responsivity and detectivity both as compared to bare ZnO NRs.

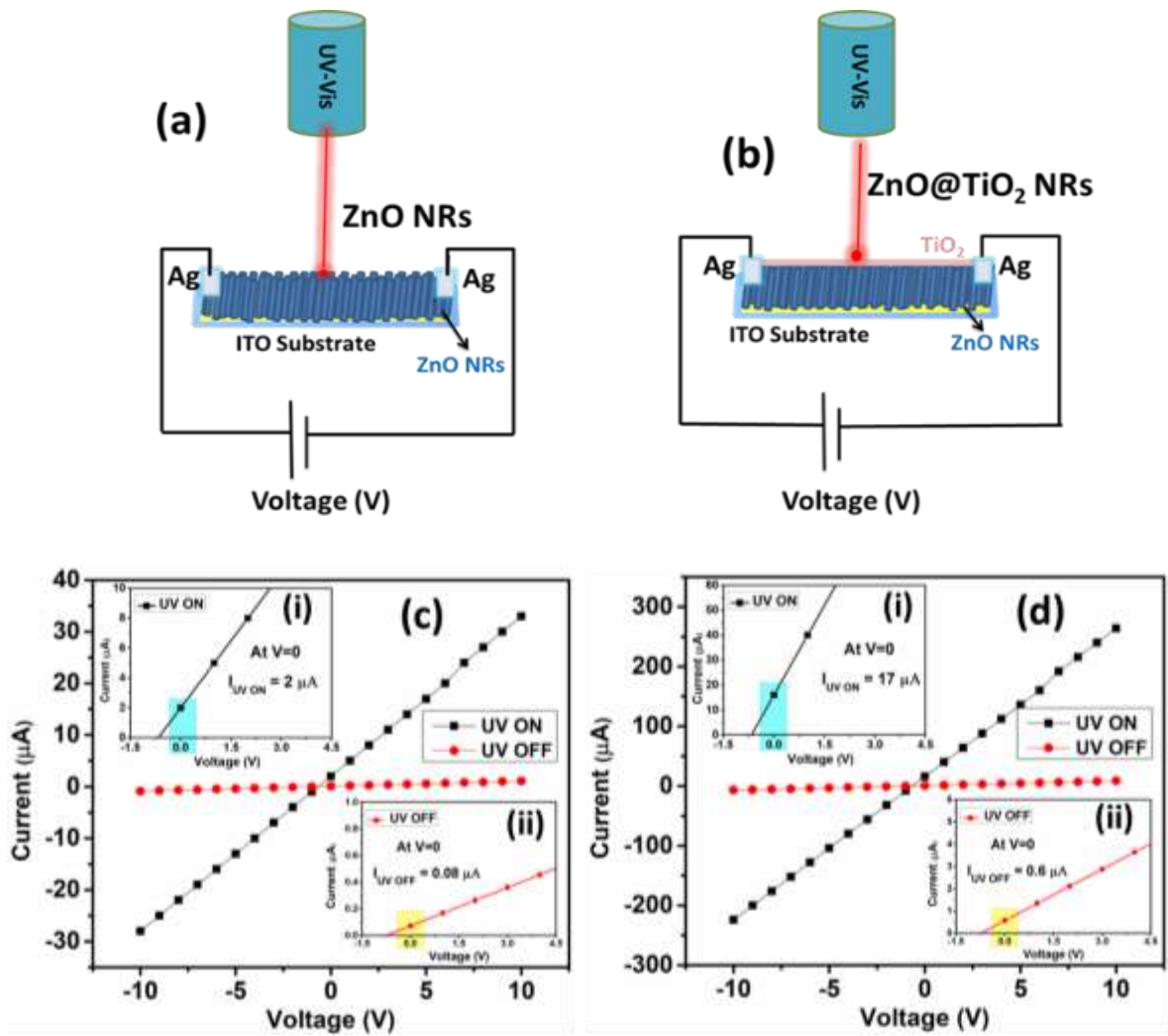

*Figure 7:* (a)-(b) Schematic of measurement setup for as prepared composite NRs using semiconducting oxides on ITO glass substrate (mechanism of producing photocurrent with applied field). Current-voltage characteristic of (c) ZnO NRs and (d) ZnO@TiO$_2$ NRs in presence of UV light, while insets of figures (c) & (d) show enlarged views of I-V characteristics.

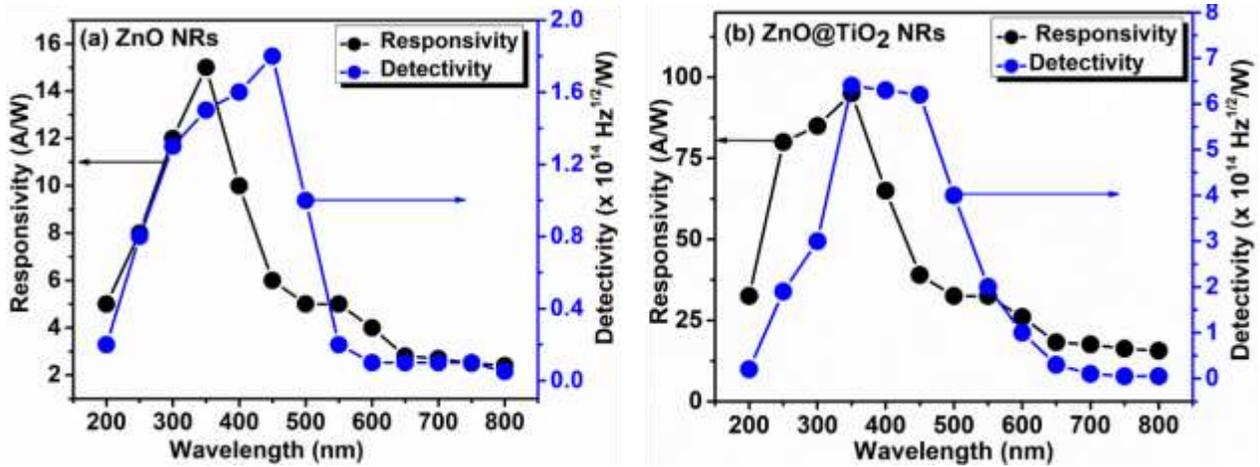

*Figure 8: Responsivity and Detectivity of (a) bare ZnO and (b) TiO$_2$ coated ZnO NRs at different wavelengths.*

In order to probe the stability of our devices, UV photocurrents of bare ZnO NRs and ZnO@TiO$_2$ NRs, at a constant bias voltage of 5 V, were measured under ambient conditions. For the purpose, the photons of wave lengths in the range 250nm - 750nm were alternatively switched ''ON'' and ''OFF'' for 10 minute each, and the results of our measurements are shown in Figs. 9 (a) and 9(b). This growth and decay of current can be used to sense UV as well as visible wavelengths. Interestingly, it is found that in case of visible light, photo current response is very low while in case of UV light photocurrent response increases ten times for prepared device. Cycles are similar and repeatable in nature, thus, confirming the reliability of both types of devices. Increase in the photocurrent under UV light is due to the promotion of a large number of electrons from the valence band to the conduction band because the photon energy exceeds the band gaps of our prepared devices. However, photons of larger wave lengths, i.e., those in the visible range, do not have sufficient energy to transfer the electrons from the valence to the conduction band, leading to considerable drop in the number of charge carriers, and, thus, the photocurrent. The photocurrent in the visible region is mainly because of the charge carriers generated due to native defects and impurities in the devices. Native defects such vacancies, self-interstitials, and anti-sites are inescapable during the synthesis of crystal lattices and have a significant impact on the performance of semiconducting oxide-based devices [42,44-47]. Significant changes in device performance, such as electronic bandgap, photocurrent, response time, etc., can be seen with a nominal change in defect concentration. This photocurrent detected in the visible area may be caused by the intrinsic oxygen deficiency found mostly in semiconducting oxides-based devices.

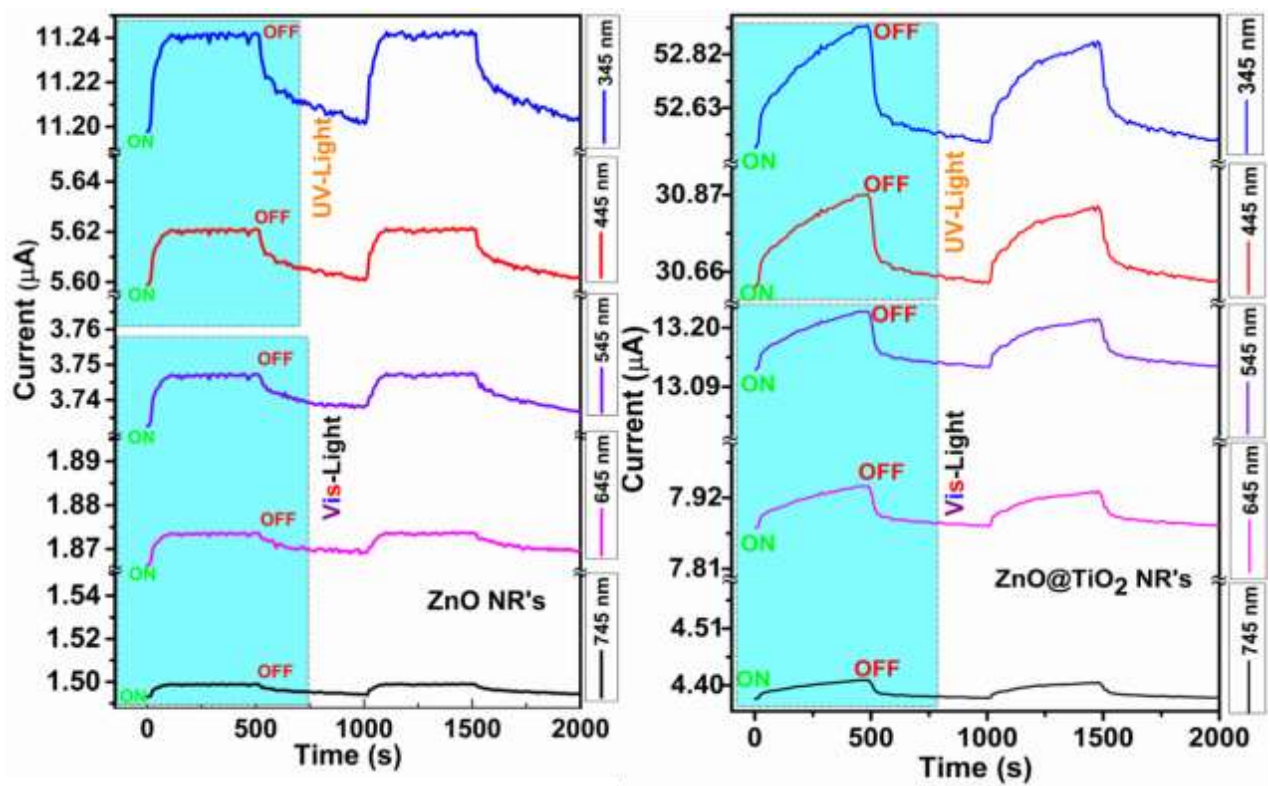

**FIGURE 9:** *Photocurrent response of (a) ZnO NRs and (b) ZnO@TiO$_2$ NRs. In both case photocurrent response is maximum with UV light, while photocurrent decreases with increase in wavelength (i.e. for Visible light).*

From Figs. 7, 8 and 9 it is obvious that ZnO@TiO$_2$ NRs has a significantly larger photocurrent response as compared to bare ZnO NRs, with a many-fold enhancement of current in the UV region. In order to systematically investigate the photo-response of the two types of devices as a function of the wavelength of the incident light, we performed measurements of the photocurrents of as prepared bare ZnO and ZnO@TiO$_2$ NRs by varying the wavelength and the results are presented in Fig. 10. From the figure it is obvious that: (a) in the entire region of wavelength probed in our experiment, the photocurrents measured in ZnO@TiO$_2$ NRs based devices is larger than that of ZnO NR based devices, and (b) the difference in two responses becomes large with the decreasing wavelengths, and for the shortest wavelength the photocurrent in the ZnO@TiO$_2$ NRs devices (53 µA) is almost five times of that in the bare ZnO devices (11.24 µA). Hence it is clearly seen that with coating of TiO$_2$ on ZnO NRs not only increases the photocurrent but also increases photosensitivity and switching stability of prepared device as compare to bare ZnO NRs.

Photocurrent in the visible region is mainly because of the charge carriers generated due to native defects and impurities in the devices. From the figure it is obvious that: (a) in the entire region of wavelength probed in our experiment, the photocurrents measured in ZnO@TiO$_2$ NRs based devices is larger than that of ZnO NR based devices, and (b) the difference in two responses becomes large with the decreasing wavelengths, and for the shortest wavelength the photocurrent in the ZnO@TiO$_2$ NRs devices (53 µA) is almost five times of that in the bare ZnO devices (11.24 µA). Hence it is clearly seen that with coating of TiO$_2$ on ZnO NRs not only increases the photocurrent but also increases photosensitivity and switching stability of prepared device as compare to bare ZnO NRs.

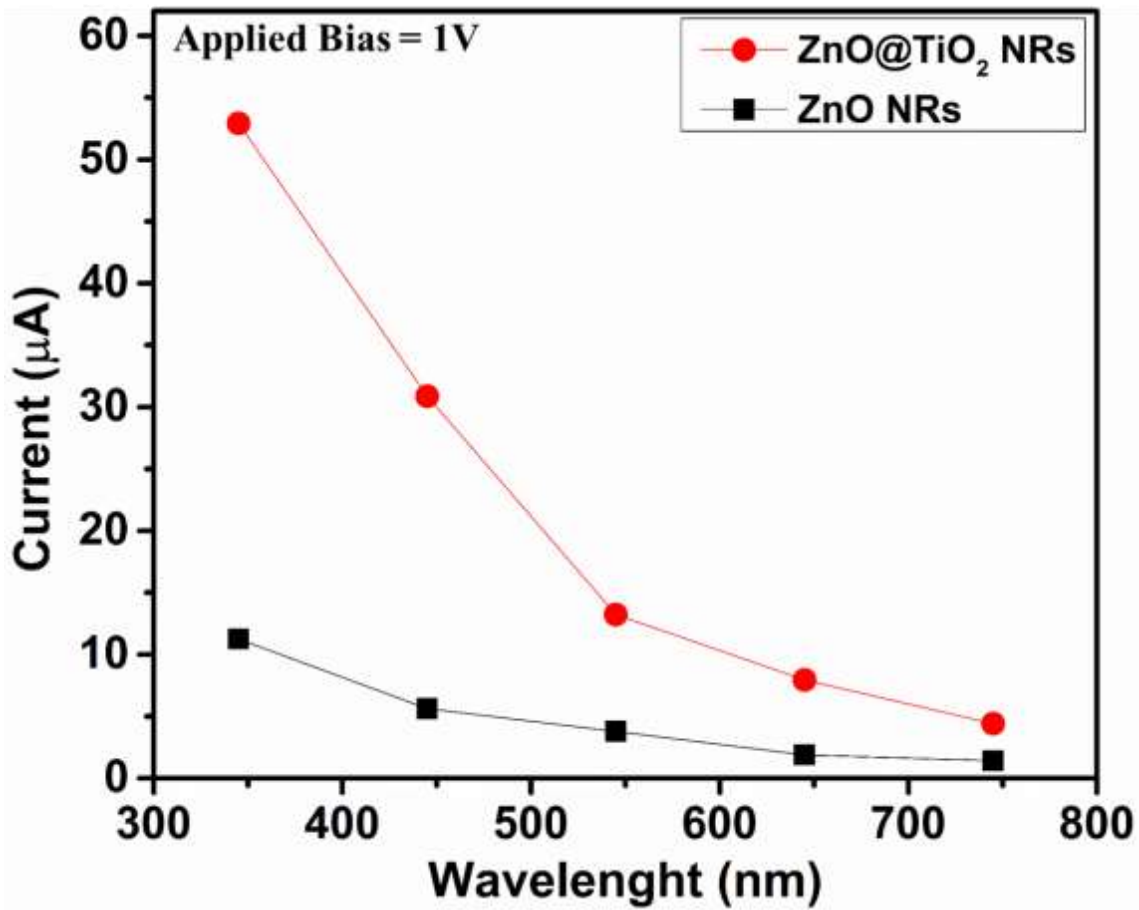

**FIGURE 10:** *Current vs wavelength plot shows photocurrent decreases with increase in the wavelength of the incident light on bare ZnO NRs and ZnO@TiO$_2$ NRs with applied bias of 1 V. ZnO@TiO$_2$ shows significantly larger photocurrent response as compared to bare ZnO NRs.*

Moreover, a comparison of the performance metrics of ZnO-based photodetectors [48-50] has been summarized in Table 1. Therefore, we believe that ZnO NRs coated with $TiO_2$ appear as one of the highly-sensitive self-powered ultraviolet photo detectors.

Table 1: Comparison of performance against other ZnO@$TiO_2$-based photodetectors.

| Photo-detectors | R (mA/W) | D ($Hz^{1/2}$/W) | Response Time | ON/OFF Ratio | Ref. |
|---|---|---|---|---|---|
| $TiO_2$/ZnO | 514 | 3.2 x $10^9$ | 33.7/12 s | 305 | [48] |
| ZnO/$TiO_2$ | 540 | 1.1 x $10^{10}$ | 9/20 s | 388 | [48] |
| Y-ZnO/$TiO_2$ NWs-Au | 27 | 6.2 x $10^{10}$ | 30 s | 1786 | [49] |
| ZnO–$TiO_2$/Si | ~4500 | - | - | 1122 | [50] |
| ZnO NRs | 15 | 1.8 x $10^{14}$ | < 30 s | 375 | **This Work** |
| ZnO@$TiO_2$ | 90 | 6.4 x $10^{14}$ | < 10s | 416 | |

**Conclusion**

In summary, photosensitive devices based on bare ZnO NRs, as well as those coated with $TiO_2$ (ZnO@$TiO_2$), were successfully fabricated. SEM analysis and X-ray diffraction confirm the morphology and structural stability of as-prepared NRs. The optical band gap was measured and a comparative blue shift was observed in ZnO@$TiO_2$ NRs. First-principles DFT calculations at the GGA+U level of theory were performed to understand the geometry, electronic structure, and optical properties of both bare as well $TiO_2$ coated NRs. Our DFT calculations, coupled with suitable parameters, correctly predict the slight blue-shift in the band gap for the ZnO@$TiO_2$ NRs, as compared to the bare ones. In addition to the band-gap shift, strong enhancements in the photoconductivity were observed ZnO@$TiO_2$, as compared to bare ZnO NRs. With the UV light falling on the sample ("ON" state), there is a significant increase in photocurrent, while in the case of no incident UV light ("OFF" state), the dark current ($I_{dark}$), decreases in nanocomposites. Responsivity and detectivity of $TiO_2$ coated ZnO NRs based device found maximum in UV region than bare ZnO NRs. ZnO@TiO$_2$ NRs show significant growth and decay in photocurrent for different wavelengths, leading to their increased photo-detection sensitivity and switching stability in the UV region, as compared to the bare ZnO NRs. Therefore, we believe that ZnO NRs coated with TiO$_2$ will prove to be very useful in fabrication of highly-sensitive self-powered ultraviolet photo detectors.

**Acknowledgments**

One of the authors (SP) acknowledges the Homi Bhabha Research Cum Teaching Fellowship (A.K.T.U.), Lucknow, India for providing financial support Through Teaching Assistantship. Authors sincerely thank Dr. Tejendra Dixit (Assistant Prof. IIIT kancheepuram) for his help during synthesis. One of the authors (SP) likes to thank Dr. Shivendra Pandey (Assistant Prof. NIT Silchar) for optical characterizations and his valuable suggestions in manuscript.

**Compliance with ethical standards:**

Conflict of interest: The authors declare that they do not have any conflict of interest.

**Data availability statement –**

The raw/processed data required to reproduce these findings cannot be shared at this time as the data also forms part of an ongoing study.